\documentclass[twocolumn,showpacs,amsmath,amssymb,pre,10pt]{revtex4}
\usepackage{amssymb}
\usepackage[final]{graphicx}

\usepackage{epstopdf}

\begin{document}

\title{Mapping Colossal Magnetoresistance Phase Transitions with the Charge-Carrier Density Collapse Model}
\author{D.O.J. Green}

\affiliation{Department of Physics, Loughborough University,
Loughborough, LE11 3TU, United Kingdom\\}

\begin{abstract}
We explain the observed specific heat anomaly (and hence entropy change) in the colossal magnetoresistive manganite $\text{Sm}_{0.55}\text{Sr}_{0.45}\text{MnO}_{3}$, by introducing phase separation into the current carrier density collapse theory via the notion of the ferromagnetic volume fraction. Within the same framework, we have also been able to explain the observed electrical resistivity of $\text{Sm}_{0.55}\text{Sr}_{0.45}\text{MnO}_{3}$ by using appropriate expressions governing the scattering mechanisms far away from the transition. Fitting specific heat and resistivity results has allowed us to separate the hopping activation energy of polaronic carriers and the bipolaron binding energy contributions to the exponential behaviour of resistivity in the paramagnetic phase.
\end{abstract}


\maketitle

\section{Introduction}

The existence of a ferromagnetic-paramagnetic and metal-insulator transition in lanthanum manganites was established in
the early 1950's \cite{jon, vans} and was extensively studied thereafter. The magnetic transition which is associated
with unusual transport properties, including large negative magnetoresistance, is observed in a family of doped manganites of perovskite structure, with the chemical
formula $\text{Re}_{1-x}\text{A}_{x}\text{MnO}_{3}$, where Re is rare earth (La, Pr, Nd, Sm...), and A is a divalent
metal (Ca, Sr, Ba...).
The recent resurgence of interest in these systems is related to the demonstration of a very large
negative magnetoresistance in thin films \cite{vonH, jin}, termed colossal magnetoresistance (CMR), which shows prospects of possible technological applications.

The metal-insulator transition in manganites has traditionally been attributed to the double exchange mechanism, which results in a varying bandwidth of electrons in the $\text{Mn}^{3+}$ d shell as a function of temperature and doping level\cite{DE, DEpapers, eff DE}. More recently it has been realized \cite{millis, millis2} that the effective carrier-spin exchange interaction of the double-exchange model is too weak to lead to a significant reduction of the electron bandwidth. Millis et al\cite{millis} proposed that there is in addition to the double exchange mechanism, a strong electron-phonon interaction arising from the Jahn-Teller splitting of the outer Mn d level, that plays a crucial role in the physics of manganites. X-ray-absorption fine structure of $\text{La}_{2/3}\text{Ca}_{1/3}\text{MnO}_{3}$ reveals differences between the nearest-neighbour Mn-O bond distributions at 80 K and above T$_{\text{c}}$. The single-site distribution at low temperature becomes a complex, multi-site distribution above T$_{\text{c}}$. The observed change is consistent with proposed small polaron related local Jahn-Teller distortions for T$>$T$_{\text{c}}$ \cite{bishop}. A giant isotope effect \cite{zhao, Oiso} suggest a very large electron phonon interaction, the sign anomaly of the Hall effect and the Arrhenious behaviour of the drift and hall mobilities, and the fact that polaron hopping satisfactorily accounts for resistivity in the paramagnetic phase \cite{Heff}, verify the polaronic nature of the charge carriers in perovskite manganites.

 Low-temperature optical \cite{optical, optical2, optical3}, electron energy-loss (EELS) \cite{EELS, Krishnan}, photoemission \cite{dessau, chuang}, and thermoelectric \cite{zhou} measurements showed that the ferromagnetic phase of manganites is not a conventional metal and confirmed that manganites were, in fact, charge-transfer doped insulators. EELS\cite{EELS, Krishnan} and O 1$_{s}$ x-ray absorption spectroscopy \cite{saitoh} consistently show oxygen p holes as current carriers as opposed to d $\text{Mn}^{3+}$ electrons proposed in the double exchange theory. Moreover Hall effect measurements by Matl et al \cite{matl} on $\text{La}_{1-x}\text{Ca}_{x}\text{MnO}_{3}$ have shown the carrier mobility near T$_{\text{C}}$ to be field independent. Also there are known classes of materials that exhibit colossal magnetoresistance in which it is guaranteed that the double exchange mechanism is non-existent such as pyrochlore manganites \cite{Shimakawa} and chromium spinels \cite{spinel}. Like the perovskite manganites these compounds exhibit very large reductions in their electrical resistivity upon application of a magnetic field. However, unlike the perovskites there is no possibility for mixed valency, a prerequisite for the double exchange mechanism \cite{ramirez}. These observations strongly suggest that the double exchange mechanism is not the origin of colossal magnetoresistance.

These observations led to a novel theory of the ferromagnetic-paramagnetic phase transition driven by non-degenerate polarons in doped charge-transfer magnetic insulators, the so-called current-carrier density collapse (CCDC) theory \cite{AS}. The CCDC theory takes into account the tendency of polarons to form local bound pairs (bipolarons) as well as the exchange interaction of p polaronic holes with d electrons. These bipolarons
are practically immobile in manganites because of the strong electron-phonon interaction, in contrast with the case for cuprates, where bipolarons are mobile and responsible for in-plane transport \cite{ABM}. The argument behind the CCDC theory is that in the paramagnetic phase a large fraction of polarons are bound into immobile bipolarons. As temperature decreases within the paramagnetic phase, the density of bipolarons increases, resulting in fewer mobile polarons, and hence, the resistivity quickly increases, reaching a maximum at $\text{T}_{\text{c}}$.  At some temperature $\text{T}_{\text{c}}$ non-degenerate polarons polarize the manganese spins via the p-d exchange interaction, $\text{J}_{\text{pd}}$, one of the polaron sub-bands falls bellow the energy of the bipolarons, thus causing pairs to break up, and the density of charge carriers (mobile polarons) increases suddenly. The occurrence of a deep minimum in mobile polaron density, near $\text{T}_{\text{c}}$ is the cause of the peak in resistivity, application of an external magnetic field will also cause a break up of bipolarons resulting in large negative magnetoresistance. The CCDC theory can explain physics of all colossal magnetoresistive materials including the pyrochlore manganites and chromium spinels.

 Hall data by Westerburg et al \cite{Westerburg} has indeed shown a current carrier density collapse. Westerburg et al \cite{Westerburg} analysed the Hall data of $\text{La}_{0.66}\text{Ca}_{0.33}\text{MnO}_{3}$ and $\text{La}_{0.66}\text{Sr}_{0.33}\text{MnO}_{3}$ it was shown in both samples that with increasing temperature there is a sharp drop in the carrier density near the metal-insulator transition, further increase of temperature causes an increase in the density of carriers, such an observation is well explained by the CCCDC theory. A similar drop in the carrier density at the metal insulator transition has been observed in $\text{Nd}_{0.5}\text{Sr}_{0.5}\text{Mn}O_{3}$ \cite{wagner}. There is other independent experimental evidence for the existence of bipolarons in perovskite manganites. Thermoelectric power measurements by Zhao et al \cite{Zhao et al}, show an isotope effect. If small polarons are bound to impurity centres, there would be no isotope effect in thermoelectric power measurements, whereas, if small polarons are bound into localized bipolaron states one would observe such an isotope effect, as the bipolaron binding energy is dependent upon the mass of isotopes. Perring et al \cite{Perring} have found short range antiferromagnetic order above T$_{C}$ in a so-called double perovskite, which may be evidence for the presence of singlet bipolarons. 
 
 A combination of low-energy electron diffraction and angle-resolved photoemission spectroscopy on some semiconducting interfaces has provided unambiguous evidence of a bipolaronic insulating state \cite{bipolaronic insulator, prb bipolaronic insulator}. There is also evidence to support the existence of immobile spin-singlet bipolarons in barium titanate perovskites \cite{bipolaron in perovskites, singlet bipolaron in perovskites} which are of a very similar structure to the CMR perovskite manganites. Furthermore x-ray and neutron scattering measurements directly demonstrate the existence of short-range correlations of polarons in the paramagnetic phase of colossal magnetoresistive perovskite manganites \cite{charge melting and polaron collapse, short range polaron correlations, correlated polarons, charge correlations}. The polaron correlations are shown to have correlation lengths of approximately 10\AA  \cite{charge correlations}. These short-range polaron correlations are shown to grow with decreasing temperature, but disappear abruptly at the ferromagnetic transition \cite{charge melting and polaron collapse, short range polaron correlations, charge correlations}. It is established that the temperature dependence of the polaron correlations is intimately related to the transport properties of the manganites \cite{short range polaron correlations, charge correlations}. Also the polaron correlations collapse under an applied magnetic field \cite{charge melting and polaron collapse, charge correlations}. The polaron correlations are interpreted as bipolarons by Nelson et al. \cite{correlated polarons}. The experimental observations of polaron correlations in the paramagnetic phase of perovskite manganites are fully consistent with the formation of bipolarons within the CCDC theory.

  Chipara et al \cite{esr measurements} claim that bipolarons are absent in, $\text{La}_{0.65}\text{Ca}_{0.35}\text{MnO}_{3}$ \cite{esr measurements} as analysis of ESR measurements has not shown the presence of bipolarons. However Chipara et al \cite{esr measurements} also accept that low-spin bipolarons (singlets) have a zero spin and, accordingly, cannot be directly inferred from ESR data. We assume, as is usually the case, that the triplet states of bipolarons lie higher in energy than the singlet state, we therefore neglect the case of triplet bipolarons. The argument of Chipara et al \cite{esr measurements} regarding the absence of bipolarons, therefore becomes irrelevant.
  
   While there is somewhat of a lack of spectroscopic evidence for the presence of bipolarons in perovskite manganites,  given all of the experimental evidence regarding the observed carrier density collapse and correlated polarons that are observed only above T$_{\text{c}}$ the idea of such a bipolaronic insulating state in perovskites does not seem unreasonable.  We believe that the CCDC theory is a much more complete theory of the colossal magnetoresistance phenomena, and other properties of these materials, when compared with the double exchange model. In this present paper we aim to show how the microscopic CCDC theory may describe the experimental specific heat and resistivity measurements in manganites, when used in conjunction with a generic phase separation model.

\section{CCDC Model and the Introduction of Phase Separation}
Alexandrov and Bratkovsky \cite{AS} have proposed that the physics of perovskite manganites including the colossal magnetoresistance effect, can be explained by the current carrier density collapse theory (CCDC).

The grand thermodynamic potential of such a system has been calculated
and has contributions from  polarons, bipolarons, localized $\text{Mn}^{+3}$ spins, and the double-counting term, respectively\cite{AS, as paper};

\begin{equation}
\Omega=\Omega_{\text{P}}+\Omega_{\text{BP}}+\Omega_{\text{S}}+\frac{1}{2}\text{J}_{\text{pd}}\text{S}\sigma \text{m}
\end{equation}

Normalizing the grand thermodynamic potential of the system by using the dimensionless temperature
$\text{t}=2k_{B}\text{T}/J_{pd}\text{S}$, magnetic field $h=2 \mu_{B}\text{H}/J_{pd}\text{S}$ and  bipolaron binding energy $\delta=\Delta/J_{pd}\text{S}$, components of the grand thermodynamic potential are found as\cite{AS, as paper};

\begin{equation}
\Omega_{\text{P}}=-2\nu \text{ty}\text{cosh}\left[\frac{\sigma+\text{h}}{\text{t}}\right]
\end{equation}

\begin{equation}
\Omega_{\text{BP}}= -\text{t}\ \text{ln}[1+\nu^{2}y^{2}\text{D}e^{2\delta/\text{t}}]
\label{Bipolaron potential}
\end{equation}

\begin{equation}
\Omega_{\text{S}}=-\text{t}\ \text{ln}\left[\frac{\text{sinh}[(5/2)(\text{m}+4\text{h})/2\text{t}]}{\text{sinh}[(1/2)(\text{m}+4\text{h})/2\text{t}]}\right]
\end{equation}

Here we have taken the $\text{Mn}^{+3}$ spin, $\text{S}=2$ due to strong Hund's coupling and the degeneracy factor, $\nu$ is 3 due to the p-character of the polarons.
If the bound pairs are extremely local objects, i.e. two holes on the same oxygen, then
they will form a singlet. However, if these holes are localized on different oxygens, then they
may well have parallel spins and form a triplet state. The triplet state is separated from the singlet state by an exchange energy, $\text{J}_{\text{st}}$.We shall
consider the simple case in which the separation of the triplets from the singlets, $\text{J}_{\text{st}}$ is
much larger than the critical temperature. In this case D=1.

From the grand thermodynamic potential a series of mean field equations describing polaron density, chemical potential, manganeese magnetization, and polaron magnetization, n, $\mu$, $\sigma$ and m, respectively has been found \cite{AS};

\begin{equation}
\text{n}=2\nu \text{y}\text{cosh}\left[\frac{\sigma+\text{h}}{\text{t}}\right]
\label{n}
\end{equation}

\begin{equation}
\text{y}^{2} =\frac{(\text{x}-\text{n})}{2\nu^{2}}e^{-2\delta/\text{t}}\
\label{ysquared}
\end{equation}

\begin{equation}
\text{m}=\text{n}\text{tanh}\left[\frac{\sigma+\text{h}}{\text{t}}\right]
\end{equation}

\begin{equation}
\sigma=\text{B}_{2}(\text{m}+4\text{h})/2\text{t})
\end{equation}
where $\text{y}=e^{\mu/\text{t}}$
The polaron density can be found as a function of $\sigma$ and temperature t by forming a quadratic equation by substituting Eq(\ref{ysquared}) into Eq(\ref{n}) . However one is left with a transcendental function to solve for $\sigma$, to overcome this one may utilize the fact that the transition is first order in a homogeneous system in a wide range of $\delta$. One can assume that the magnetization behaves as a step function that can become ``smeared'' due to disorder.
Alexandrov et al \cite{coex} suggested that one should consider that different parts of a particular sample will exhibit differing values of $\delta$ and therefore have different transition temperatures, and that a Gaussian distribution would provide a reasonable model for the spread of these transition temperatures. One can therefore say that the normalized magnetization (which we call the volume fraction, V) is equal to 1 minus the cumulative distribution function of transition temperatures. The volume fraction is then given by;

\begin{equation}
\text{V}=\frac{1}{2}\text{erfc}\left[\frac{\text{t}-\text{t}_{\text{c}}}{\Gamma}\right]
\label{volfractioneq}
\end{equation}
$\Gamma= s\sqrt{2}$, where s is the standard distribution of transition temperatures across the sample.

We term the normalized magnetization as the volume fraction as it describes the fraction of the sample that remains in the ferromagnetic phase as the temperature is increased.

Broadening of the magnetic transition is in agreement with numerical studies by Mercaldo et al \cite{mercaldo} who have shown that introducing quenched disorder, such as grain and crystalline boundaries into a three-dimensional pure system that exhibits a first order transition causes broadening of the temperature dependence of magnetization. Bulk samples of perovskite manganites usually exhibit crystallite size in the order of nanometers\cite{MEg, raita, nano}, thus leading to nanoscale phase separation in manganites \cite{phasesepbook}.

\section{Transition Temperature and Volume Fraction of $\text{Sm}_{0.55}\text{Sr}_{0.45}\text{MnO}_{3}$}

\begin{figure}[!ht]
\centering
\includegraphics[scale=0.65]{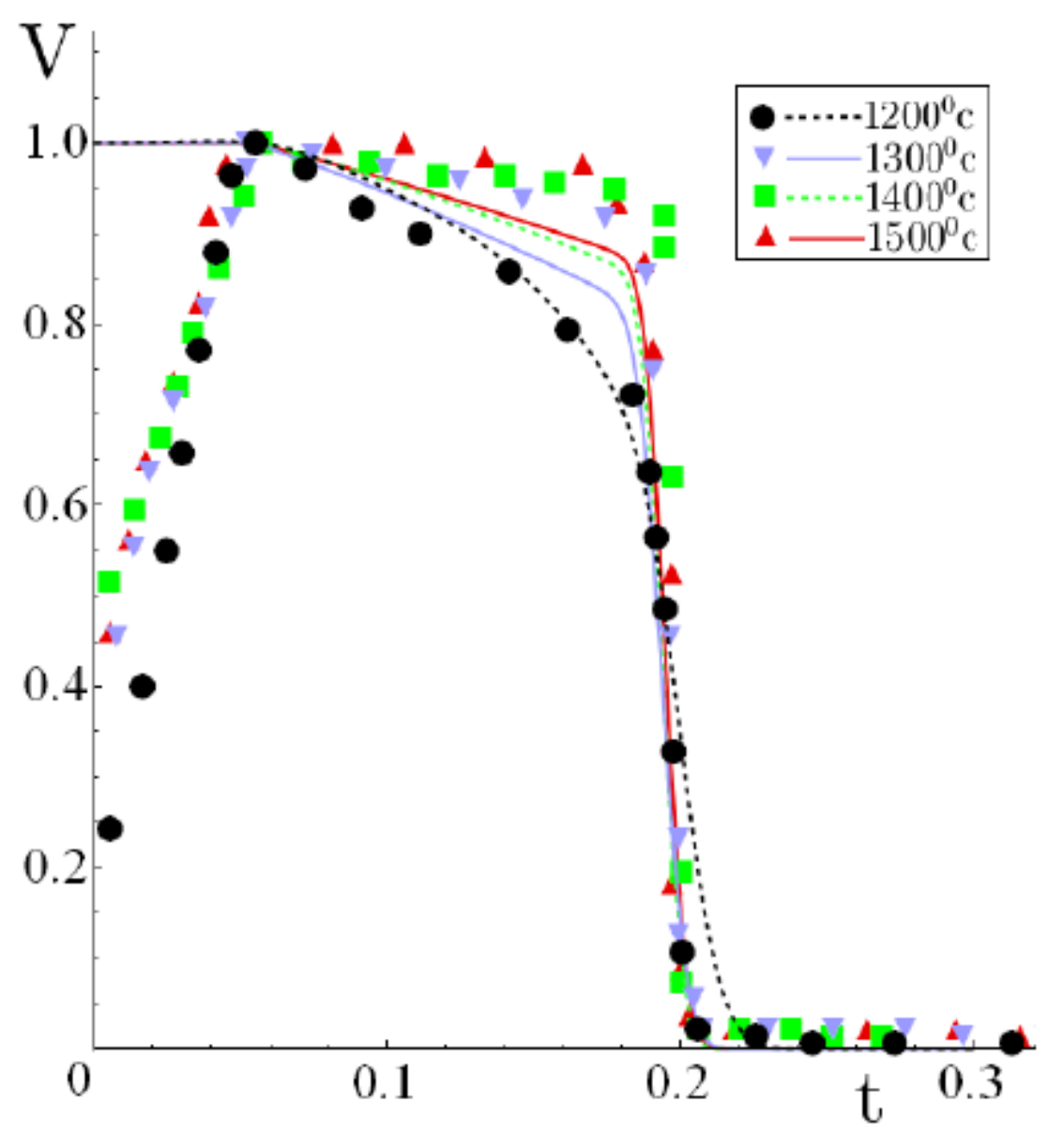}
\caption{A graph of the volume fraction of $\text{Sm}_{0.55}\text{Sr}_{0.45}\text{MnO}_{3}$, normalized such that the value of half volume fraction occurs at approximately, $\text{t}=0.2$, as calculated for $\delta=0.65$ and $x=0.45$. Solid circles represent scaled data \cite{MEg}, lines represent expressions used to describe the volume fraction.}
\label{volfrac}
\end{figure}

Egilmez et al \cite{MEg, meg2} have produced samples of $\text{Sm}_{0.55}\text{Sr}_{0.45}\text{MnO}_{3}$ annealed at different temperatures to produce different grain and crystallite size. Annealing at higher temperatures produced larger grains and crystallites, reducing the density of grain boundaries, and hence reduced disorder. The extra disorder in the low temperature annealed samples results in broadening of the transition and hence volume fraction, see Fig(\ref{volfrac}). The erfc function, Eq(\ref{volfractioneq}) corresponds to a Gaussian distribution of transition temperatures, however the magnetization data of $\text{Sm}_{0.55}\text{Sr}_{0.45}\text{MnO}_{3}$ can be better fitted using combinations of polynomials and the erfc function. In general one can use any expression to describe the volume fraction, as in Fig(\ref{volfrac}).

\begin{figure}[!ht]
\centering
\includegraphics[scale=0.75]{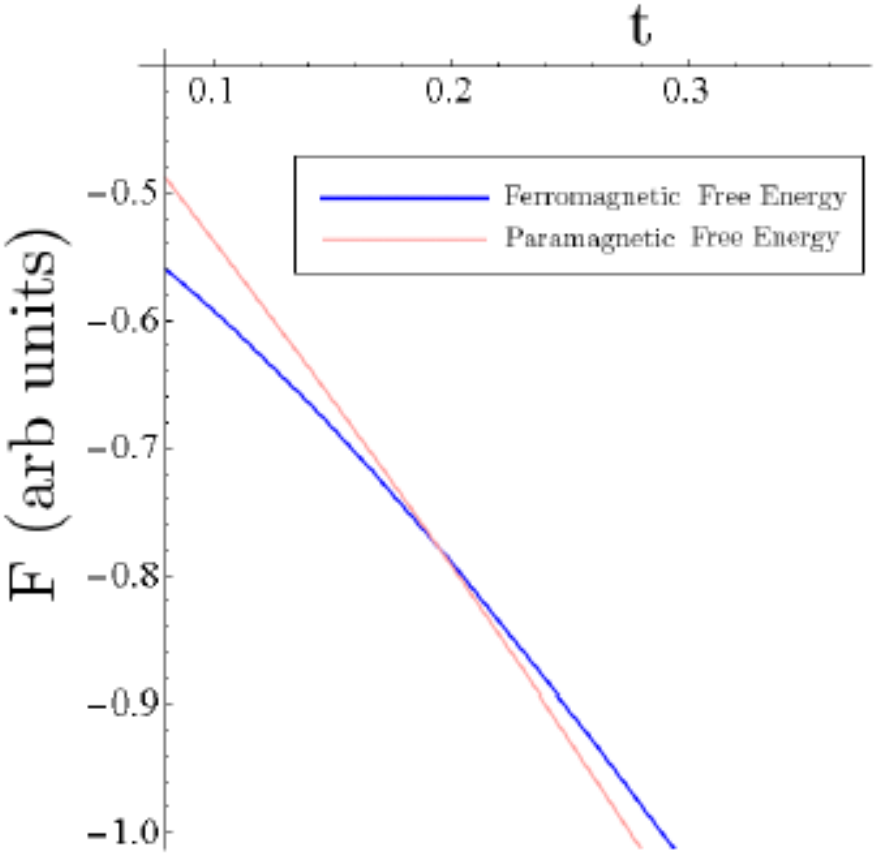}
\caption{A graph of the Helmholtz free energy, for $\delta=0.65$ and $x=0.45$, in the ferromagnetic and paramagnetic phases. The plot indicates $\text{t}_{\text{c}}\approx0.2$.}
\label{ssmofreeenergy}
\end{figure}

One should remember that within the CCDC model that although the density of polarons is temperature and field dependent the total number of holes doped into the system is constant and equal to the doping concentration, N$=$x.
Utilizing the fact that Helmholtz free energy is continuous over a phase transition we may calculate, $\text{F}=\Omega+\mu \text{N}$ in both the ferromagnetic and paramagnetic phases, by substituting $\sigma=1\  \text{or}\ 0$ respectively into the expressions for $\Omega$, and $\mu$.
The temperature at which these two functions cross over is the transition temperature. Fig(\ref{ssmofreeenergy}) shows a plot of free energy in the CCDC model for both phases, with $\delta=0.65$ and x$=0.45$. This value of $\delta$ was chosen such that a reasonable agreement between experimental specific heat data of $\text{Sm}_{0.55}\text{Sr}_{0.45}\text{MnO}_{3}$ \cite{MEg, meg2} and CCDC theory can be achieved (section(IV)). We have assumed $\delta=0.65$ for all samples of $\text{Sm}_{0.55}\text{Sr}_{0.45}\text{MnO}_{3}$ to simplify the problem somewhat.

The normalized magnetization data for $\text{Sm}_{0.55}\text{Sr}_{0.45}\text{MnO}_{3}$ \cite{MEg}, has been scaled such that the value of half volume fraction occurs at $\text{t}=0.2$, see Fig(\ref{volfrac}).

The transition temperature $\text{T}_{\text{c}}$ in $\text{Sm}_{0.55}\text{Sr}_{0.45}\text{MnO}_{3}$ samples only varies by a few degrees Kelvin \cite{MEg}, we can therefore look at the average conversion factor between the temperature in Kelvin and the normalized temperature t, which is  $1.53\times10^{-3}$. This corresponds to an exchange energy between the carriers and localized spins of $J_{pd}=56\text{meV}$, and hence the bipolaron binding energy is found to be $\Delta= 72.8\text{meV}$ as $\delta=0.65$. The value of $J_{pd}$ found is in close agreement with the value of $60\text{meV}$ as estimated for perovskite manganites by \cite{zhao from correction}.

\section{Specific Heat and Entropy of $\text{Sm}_{0.55}\text{Sr}_{0.45}\text{MnO}_{3}$}

We can calculate the entropy of the system as:
\begin{equation}
-\left(\frac{\partial \text{F}}{\partial \text{t}}\right)_{\text{NV}}=-\left(\frac{\partial \Omega}{\partial \text{t}}\right)_{\text{NV}}-x\left(\frac{ \partial \mu}{\partial \text{t}}\right)_{\text{NV}}=\text{S}
\label{entropyfromf}
\end{equation}

In eq (\ref{entropyfromf}) $\Omega$ is the total grand thermodynamic potential found from the CCDC theory. By substituting $\sigma=1\  \text{or}\ 0$ into equation(\ref{entropyfromf}) and retaining the value of $\delta$ as 0.65, we may calculate the entropy of the ferromagnetic and paramagnetic phases respectively.

We suggest that the total entropy incorporating phase separation can be described as;

\begin{equation}
 \text{S}_{\text{tot}}=\text{VS}_{f}+(1-\text{V})\text{S}_{p}
 	\label{totalentropyequation}
\end{equation}
where $\text{S}_{f}$ and $\text{S}_{p}$ denote the entropy in the ferromagnetic and paramagnetic phases respectively, and V is the volume fraction.

Differentiating once more we may calculate the specific heat.

\begin{equation}
 \text{C}_{\text{v}}=\frac{\partial \text{S}_{\text{tot}}}{\partial t}
 	\label{specific heat}
\end{equation}

\begin{figure}[!ht]
\centering
\includegraphics[scale=0.78]{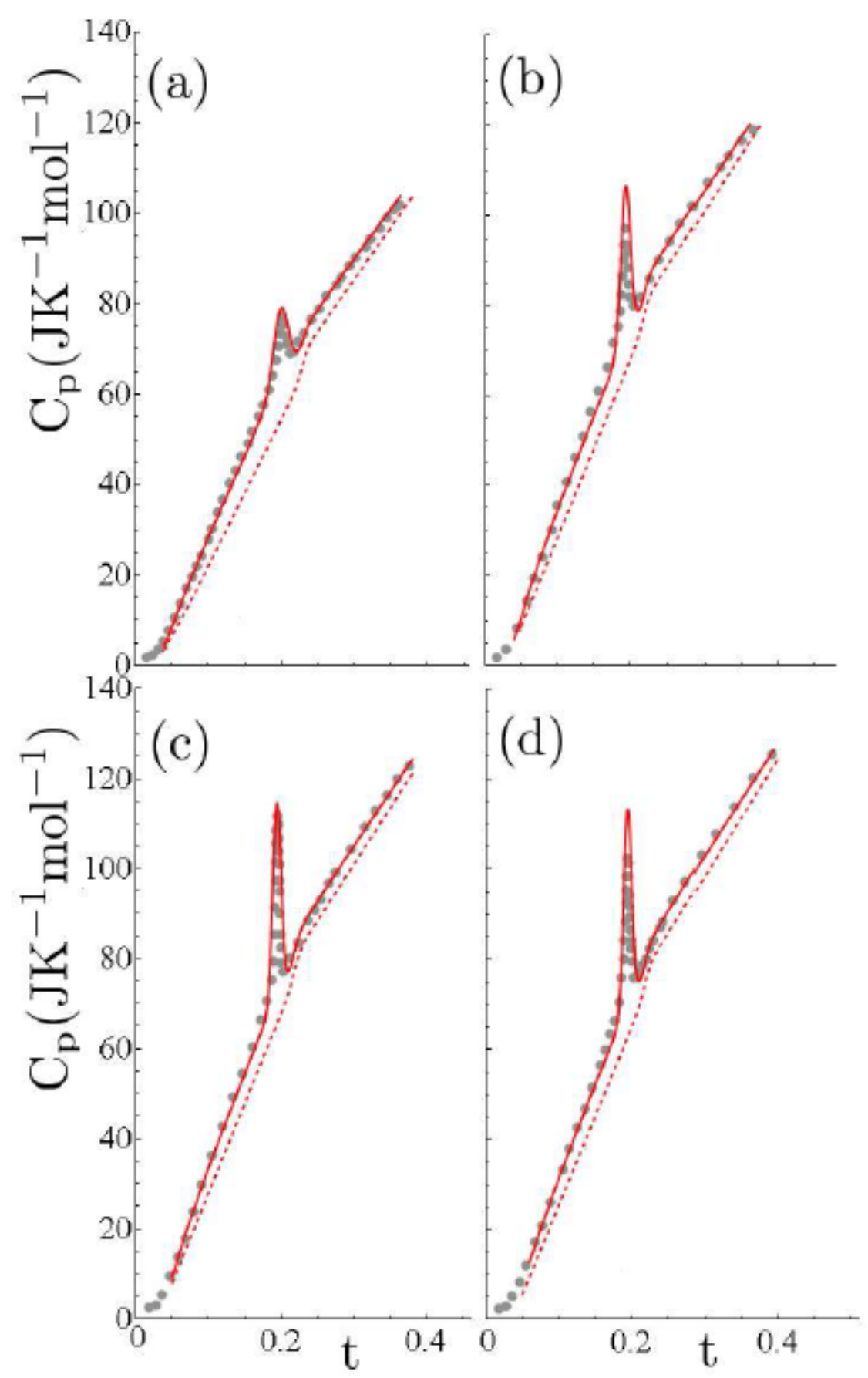}
\caption{Plots of specific heat of $\text{Sm}_{0.55}\text{Sr}_{0.45}\text{MnO}_{3}$ annealed at (a)$1200^\circ\text{C}$, (b)$1300^\circ\text{C}$, (c)$1400^\circ\text{C}$, (d)$1500^\circ\text{C}$. Solid circles represent experimental specific heat data\cite{MEg, meg2}, dashed lines correspond to our estimates of lattice specific heat, solid lines represent the addition of the lattice specific heat and the theoretical CCDC component.}
\label{total specific heat}
\end{figure}

We fit our calculated specific heat under constant volume to the experimentally measured specific heat under constant pressure of $\text{Sm}_{0.55}\text{Sr}_{0.45}\text{MnO}_{3}$, by assuming no thermal expansion.
To fit to the experimental data an estimate of the lattice specific heat is required, typically the phonon blank method is used. The phonon blank method requires another sample to be prepared that does not exhibit the transition in question, this sample should have the same crystal structure and as close as possible a similar atomic mass. A possible phonon blank material for $\text{Sm}_{0.55}\text{Sr}_{0.45}\text{MnO}_{3}$ could be produced by replacing Mn for Ti atoms, the resultant material would not exhibit the ferromagnetic-paramagnetic transition, if this material possessed the same crystal structure as $\text{Sm}_{0.55}\text{Sr}_{0.45}\text{MnO}_{3}$ the specific heat of that sample would be very close to the lattice specific heat of $\text{Sm}_{0.55}\text{Sr}_{0.45}\text{MnO}_{3}$. However no phonon blanks for $\text{Sm}_{0.55}\text{Sr}_{0.45}\text{MnO}_{3}$ were produced. In fact we are unaware of any published specific heat data for perovskite manganites including a phonon blank. We therefore simply take an estimate of the smooth lattice background specific heat, see Fig(\ref{total specific heat}).

\begin{figure}[!ht]
\centering
\includegraphics[scale=0.78]{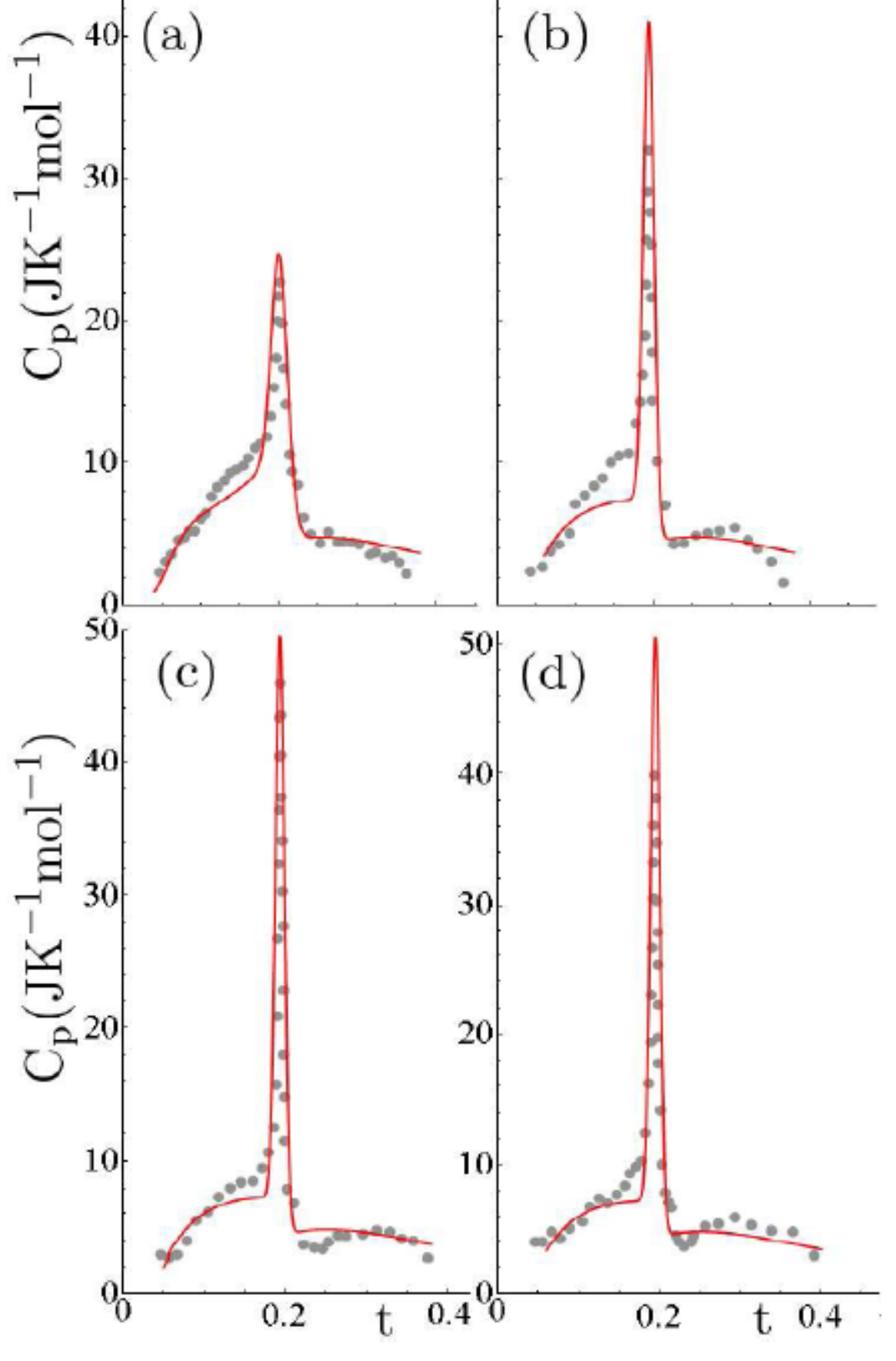}
\caption{Plots of the anomalous component of specific heat of $\text{Sm}_{0.55}\text{Sr}_{0.45}\text{MnO}_{3}$ annealed at (a)$1200^\circ\text{c}$, (b)$1300^\circ\text{c}$, (c)$1400^\circ\text{c}$, (d)$1500^\circ\text{c}$. Solid circles represent the difference between the experimental specific heat data \cite{MEg, meg2} and the estimate of the lattice specific heat, solid lines represent the specific heat calculated from the CCDC theory. }
\label{anomalous specific heat}
\end{figure}

Figs(\ref{total specific heat}) and (\ref{anomalous specific heat}) show that the theoretical specific heat is in reasonable agreement with the experimental results assuming our estimate of lattice specific heat is correct.

\begin{figure}[!ht]
\centering
\includegraphics[scale=0.75]{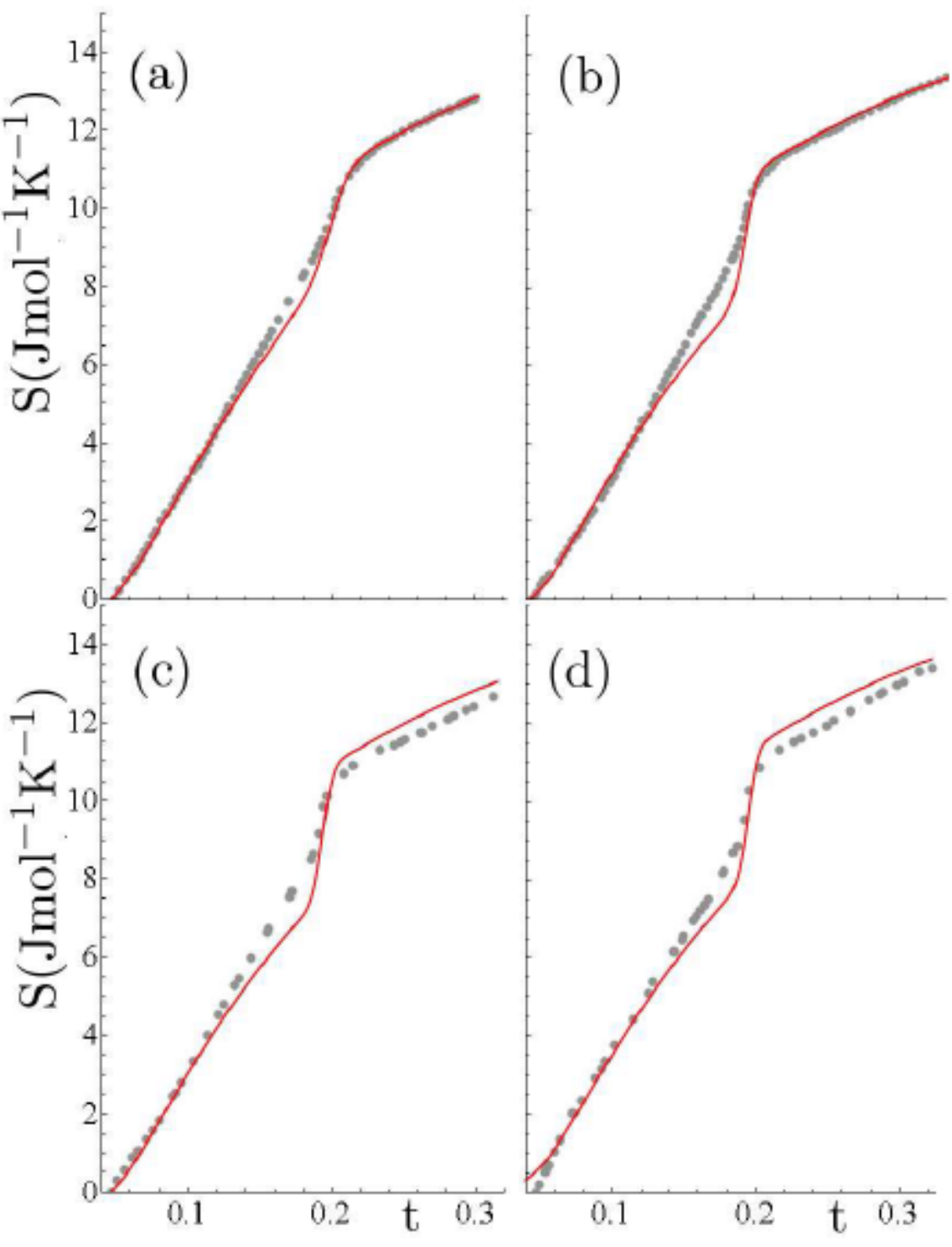}
\caption{Plots of entropy of $\text{Sm}_{0.55}\text{Sr}_{0.45}\text{MnO}_{3}$ annealed at (a)$1200^\circ\text{C}$, (b)$1300^\circ\text{C}$, (c)$1400^\circ\text{C}$, (d)$1500^\circ\text{C}$. }
\label{entropy}
\end{figure}

We can calculate the entropy change associated with the anomalous part of the experimental data,
Fig(\ref{anomalous specific heat}), by means of;

\begin{equation}
\text{S}=\int\frac{\text{\text{C}}_{\text{v}}}{\text{t}}\text{dt}
\label{cv to s}
\end{equation}
The entropy associated with the specific heat anomaly in the experimental data is plotted alongside the theoretically determined entropy in Fig(\ref{entropy}).

We investigate the entropy change from $\text{t}=0.0465$ to $\text{t}=0.3$, as these temperatures are well below and above the transition respectively, hence all samples should exhibit the same entropy change over this temperature interval. The only role that the phase separation plays is to broaden the observed specific heat peak. Table(\ref{entropy change}) shows the entropy change associated with the anomalous part of the data for each sample.

 \begin{table}[!ht]
\begin{tabular}{ |p{4.3cm} | p{0.6cm} | p{0.6cm} | p{0.6cm} | p{0.6cm} |}
\hline
T$_{\text{a}}$($^{0}\text{c}$)&1200&1300&1400&1500\\ \hline
 $\Delta \text{S}_{\text{data}}$ $\left(\text{J}\text{K}^{-1}\text{mol}^{-1}\right)$ (1.dp)&12.9&12.8&12.4&13.0\\
 \hline
 \end{tabular}
  \caption{Entropy change associated with the anomalous component of the experimental specific heat data of $\text{Sm}_{0.55}\text{Sr}_{0.45}\text{MnO}_{3}$, where T$_{a}$ is the annealing temperature. }
 \label{entropy change}
\end{table}

Averaging the data from table(\ref{entropy change}) we find the entropy change associated with the anomalous part of the experimental data from,  $\text{t}=0.0465$ to $\text{t}=0.3$ as, $\Delta \text{S}_{\text{data}}=12.8 \pm 0.1 \text{J}/\text{K}^{-1}\text{mol}^{-1}$. This is in good agreement with the theoretical entropy change over this temperature range, $\Delta \text{S}_{\text{CCDC}}=12.802\text{J}/\text{K}^{-1}\text{mol}^{-1}$.

Several authors report that the entropy change associated with the ferromagnetic-paramagnetic transition is much smaller than expected \cite{meg2,lsmo} and hence unexplainable. Tanaka and Mitsuhashi\cite{tanaka} found a very small entropy change in $\text{La}_{0.8}\text{Ca}_{0.2}\text{MnO}_{3}$, Ramirez \cite{ramirez} states that this is approximately $10\%$ of the expected value of $\approx 11\text{J}\text{K}^{-1}\text{mol}^{-1}$ (for localized spins), approximately twice as much entropy was found to reside under the ordering peak in $\text{La}_{0.67}\text{Ca}_{0.33}\text{MnO}_{3}$ by Ramirez et al \cite{ramirez96}.
 The reasons for the entropy change being left unexplained up until now are that most authors\cite{meg2,lsmo} have only considered local-moment degrees of freedom. The contribution to entropy of polarons acts to reduce the entropy change at $\text{T}_{\text{c}}$. Below $\text{T}_{\text{c}}$ polarons are unpaired, above $\text{T}_{\text{c}}$ the polarons are bound into immobile bipolarons,  and hence the entropy of the charge carriers actually decreases at $\text{T}_{\text{c}}$. The total entropy change at $\text{T}_{\text{c}}$ will always be positive, it is simply that the contribution of carriers acts to reduce this increase, resulting in entropy changes that are smaller than has been expected \cite{meg2,lsmo}.

\section{Resistivity of $\text{Sm}_{0.55}\text{Sr}_{0.45}\text{MnO}_{3}$}

Using the same expressions governing the volume fraction, we fit the experimental resistivity data of $\text{Sm}_{0.55}\text{Sr}_{0.45}\text{MnO}_{3}$ \cite{MEg, meg2}.

\begin{figure}[!ht]
\centering
\includegraphics[scale=0.68]{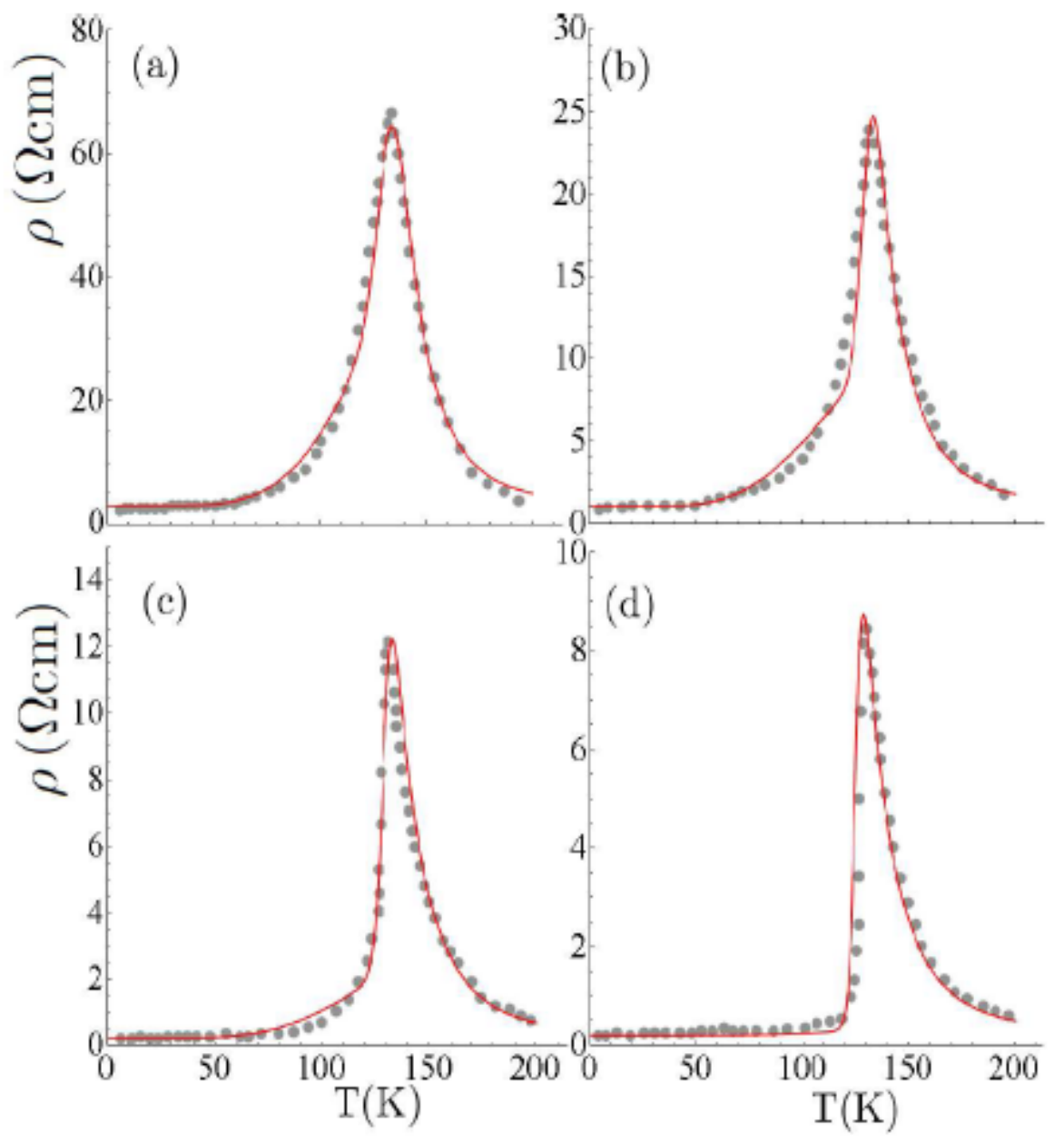}
\caption{Plots of the electrical resistivity of $\text{Sm}_{0.55}\text{Sr}_{0.45}\text{MnO}_{3}$ annealed at (a)$1200^\circ\text{C}$, (b)$1300^\circ\text{C}$, (c)$1400^\circ\text{C}$, (d)$1500^\circ\text{C}$. Solid circles show the experimental data \cite{MEg, meg2}], lines show fitted theoretical resistivity from equation(\ref{analyticalexpressionforresistivity}).}
\label{resistivity}
\end{figure}

A simple analytical expression for the resistivity of a binary mixture, which is consistent with comprehensive numerical simulations was proposed by Kabanov et al \cite{kabanov} and considered for use, when explaining the resistivity of mixed phase perovskite manganites \cite{coex};\newline

\begin{equation}
\rho_{\text{eff}}=\rho_{\text{ferro}}^{\text{V}}\rho_{\text{para}}^{1-\text{V}},
\label{analyticalexpressionforresistivity}
\end{equation}

which is valid in a wide range of the ratios $\rho_{\text{para}}/\rho_{\text{ferro}}$\cite{kabanov}, $\text{V}$ is the volume fraction. The resistivity is fitted in some-what of a generic manner, in as much that we are mixing the phases of a metal-like phase with that of an insulating phase. The expressions used are however, consistent with the physics of the CCDC model.

Here we have used the expression governing the resistivity in the ferromagnetic phase as,\begin{equation}
\rho_{ferro}=\text{a}\text{T}^{4.5} + \text{c}/\text{sinh}^{2}\left[\hbar\omega_{0}/\text{k}_{\text{B}}\text{T}\right].
\label{ferrores}
\end{equation}

In fitting to the experimental data the coefficient c is found as $1(\Omega\ \text{cm})$ for all samples and $\hbar\omega_{0}$ has been fixed as; 50meV, which is a typical value for the phonon cut off in LCMO \cite{as paper}.

The expression for the ferromagnetic resistivity used here was suggested by Zhao et al\cite{zhaotransport}. Kubo and Ohata \cite{kubo} have studied the magnon scattering for half metals where the spin-up (minority) and spin-down (majority) bands are well separated. In this case, one-magnon scattering is forbidden, and thus the two-magnon process plays a role in the low-temperature resistivity, which gives a contribution that is proportional to $\text{T}^{4.5}$. The term  $\text{c}/\text{sinh}^{2}\left[\hbar\omega/\text{k}_{\text{B}}\text{T}\right]$ comes from the relaxation time of polarons calculated by Lang and Firsov \cite{langfirsov}. Jamie et al \cite{jamie} have ruled out electron-electron, and electron-phonon scattering mechanisms in the low temperature conductivity of manganites, that would give $\text{T}^{2}$ and $\text{T}^{5}$ contributions respectively.

The resistivity of the paramagnetic phase can be well described by \cite{coex};
\begin{equation}
\rho_{para}=\text{bT} e^{\frac{E}{\text{k}_{\text{B}}\text{T}}}
\label{para res}
\end{equation}

We also include a residual resistivity $\rho_{0}$. The parameters used to fit the resistivity are shown in table(\ref{parameters}).

\begin{table}[!ht]
\begin{tabular}{ |p{3.8cm} | p{0.8cm} | p{0.6cm} | p{0.6cm} | p{0.6cm}|}
\hline
T$_{\text{a}}$ ($^\circ\text{C}$) & 1200& 1300& 1400&1500\\
\hline
$\rho_{0}$ $\left(\Omega\ \text{cm}\right)$& 3 & 1 & 0.21& 0.2\\
    \hline
   a $\left(\Omega\ \text{cm K}^{-4.5}\right)\times10^{-10}$ & 38&17 &4& 0.07\\
    \hline
      b $\left(\Omega\ \text{cm K}^{-1}\right)\times10^{-6}$&2.2&1.3 & 1.25&1.2\\
      \hline
       E $\left(\text{meV}\right)$  &138 & 131&123&116\\
        \hline
        E$_{\text{a}}$ $\left(\text{meV}\right)$  &101.6 & 94.6 &86.6&79.6\\
        \hline
       \end{tabular}
       \caption{Parameters governing the resistivity of $\text{Sm}_{0.55}\text{Sr}_{0.45}\text{MnO}_{3}$ \cite{MEg, meg2}}
       \label{parameters}
       \end{table}

 It is clear that the peaks observed in the resistivity of $\text{Sm}_{0.55}\text{Sr}_{0.45}\text{MnO}_{3}$, Fig(\ref{resistivity}), are broadened with increasing disorder and phase separation. One can also see that the resistivity is increased with the increasing disorder, this is simply due to increased scattering rates, as can be seen by the increase values of the constants a,b and $\rho_{0}$ in table(\ref{parameters}) with decreasing annealing time.

The high temperature mobility of polaronic carriers is likely to be dominated by hopping events due to polaron band narrowing, \cite{langfirsov, polaronsadvmat}, therefore E has contributions from the bipolaron binding energy, $\Delta$, and the activation energy of polarons, $\text{E}_{\text{a}}$ as;

\begin{equation}
\text{E}=\text{E}_{\text{a}}+\Delta/2
\end{equation}

By fitting the specific heat we have found $\Delta/2= 36.4\text{meV}$, and from fitting the resistivity we have found E. This has allowed us to separate the polaron hopping activation energy and bipolaron binding energy from the exponential behaviour of the resistivity in the paramagnetic regime. See table(\ref{parameters}) for values of polaron hopping activation energy.
 We can see that the hopping activation increases with an increasing density of grain boundaries (decreasing annealing temperature)\cite{MEg}. This is what one would expect to see as the increasing levels of disorder act to localize the carriers resulting in a larger activation energy required for hopping.

\section{Concluding Remarks}

Introducing Phase separation into the thermodynamics of the microscopic CCDC model by mixing the high and low temperature phases has allowed for the specific heat of $\text{Sm}_{0.55}\text{Sr}_{0.45}\text{MnO}_{3}$ samples produced by Egilmez et al \cite{MEg, meg2} to be well explained. Up until now the entropy change associated with the ferromagnetic-paramagnetic transition in colossal magnetoresistive perovskite manganites has been left largely unexplained, however we find the entropy change from the CCDC theory and the experimental data to be in close agreement. The contribution to entropy of polarons acts to reduce the entropy change at $\text{T}_{\text{c}}$. Below $\text{T}_{\text{c}}$ polarons are unpaired, above $\text{T}_{\text{c}}$ the polarons are bound into immobile bipolarons, and hence the entropy of the charge carriers actually decreases at $\text{T}_{\text{c}}$. The total entropy change at $\text{T}_{\text{c}}$ will always be positive, it is simply that the contribution of carriers acts to reduce this increase, which results in an entropy change that is smaller than has been expected \cite{meg2,lsmo}. Fitting specific heat data, in conjunction with a generic phase separated model of resistivity has allowed bipolaron binding energy and polaron hopping activation energy contributions to the exponential behaviour of resistivity to be separated. We hope that experimentalists working in the field of perovskite manganites try to produce phonon-blank materials such that the lattice contribution to specific heat of samples can be accurately determined.

We accept that the CCDC theory is not the only theory that hopes to explain the physics of colossal magnetoresistive materials, we have shown however for the first time that this theory may be used to explain the specific heat of perovskite manganites. More work is required, especially by experimentalists to either conclusively prove or disprove this model.

\end{document}